\documentclass[prl,twocolumn,twoside,showpacs,floatfix]{revtex4}
\usepackage{epsf,amssymb,mathptmx}
\begin{document}
\noindent [Phys. Rev. Lett. {\bf 91}, 148701 (2003)]
\title{Sandpile on scale-free networks}
\author{K.-I. Goh, D.-S. Lee, B. Kahng and D. Kim\\} 
\affiliation{
\mbox{School of Physics and Center for Theoretical Physics,
Seoul National University, Seoul 151-747, Korea}}
\date{\today}

\begin{abstract}
We investigate the avalanche dynamics of the Bak-Tang-Wiesenfeld
(BTW) sandpile model on scale-free (SF) networks, where threshold height
of each node is distributed heterogeneously, given as 
its own degree. We find that the avalanche size distribution 
follows a power law with an exponent $\tau$. 
Applying the theory of multiplicative branching process, we obtain the 
exponent $\tau$ and the dynamic exponent $z$ as a function 
of the degree exponent $\gamma$ of SF networks as  
$\tau=\gamma/(\gamma-1)$ and $z=(\gamma-1)/(\gamma-2)$ 
in the range $2 < \gamma < 3$ and the mean field values $\tau=1.5$ 
and $z=2.0$ for $\gamma >3$, with a logarithmic correction at $\gamma=3$.
The analytic solution supports our numerical simulation results.
We also consider the case of uniform threshold, finding that 
the two exponents reduce to the mean field ones. 
 
\end{abstract}
\maketitle
Recently the emergence of a power-law degree 
distribution in complex networks, $i.e.$, $p_d(k)\sim k^{-\gamma}$ 
with the degree exponent $\gamma$, have attracted many attentions~
\cite{rmp,adv}. 
Such networks, called scale-free (SF) networks, are ubiquitous 
in nature. Due to the heterogeneity in degree, SF networks are 
vulnerable to attack on a few nodes with large degree~\cite{bara}. 
However more severe catastrophe can occur, triggered by a small 
fraction of nodes but causing a cascade of failures of other 
nodes~\cite{watts}. 
The 1996 blackout of power transportation network in Oregon and 
Canada is a typical example of such a cascading failure~\cite{grid}. 
As another example, malfunctioning router will automatically prompt Internet 
protocols to bypass the missing router by sending packets to 
other routers. If the broken router carries 
a large amount of traffic, its absence will place a significant 
burden on its neighbors, which might bring the failure of the 
neighboring routers again, leading to a breakdown of the entire system 
eventually~\cite{motter}. 

To understand such cascading failures on SF networks, 
we study in this Letter the Bak-Tang-Wiesenfeld (BTW) sandpile 
model~\cite{btw} as 
a prototypical theoretical model exhibiting avalanche behavior. 
The main feature of the model on the Euclidean space is the 
emergence of a power law 
with exponential cutoff in the avalanche size distribution, 
\begin{equation}
p_a(s) \sim s^{-\tau}\exp(-s/s_c),
\end{equation}
where $s$ is avalanche size and $s_c$ its characteristic size.
While many studies of the BTW sandpile model and its related models 
have been carried out on the Euclidean space, the study of them 
on complex networks has rarely been carried out. 

Bonabeau~\cite{bonabeau} have studied the BTW sandpile 
model on the Erd\"os-R\'enyi (ER) random networks and found 
that the avalanche size distribution follows a power law with 
the exponent $\tau \simeq 1.5$, consistent with the mean field solution 
in the Euclidean space \cite{alstrom}. Recently Lise and 
Paczuski~\cite{paczuski} have studied the Olami-Feder-Christensen 
model~\cite{ofc} on regular ER networks, where degree 
of each node is uniform but connections are random. 
They found the exponent to be $\tau\approx1.65$. 
However, when degree of each node is not uniform, they found 
no criticality in the avalanche size distribution. 
Note that they assumed that the threshold of each node is uniform, 
whereas degree is not.
While such a few studies have been performed on ER random 
networks, the study of the BTW sandpile model on SF 
networks has not been performed yet, even though there 
are several related applications as mentioned above. 

We study the dynamics of the BTW sandpile 
model on SF networks both analytically and numerically. 
In the model, we first consider the case where threshold 
height of each node is assigned to be equal to its degree, 
so that threshold is not uniform but distributed following 
the power law of the degree distribution. 
An analytic solution for the avalanche size and duration
distributions is obtained by applying the theory of 
multiplicative branching process developed by Otter 
in 1949~\cite{otter}. 
The multiplicative branching process approach was used to 
obtain the mean-field solution for the BTW model in the Euclidean 
space~\cite{alstrom}, which is valid 
above the critical dimension $d_c=4$. 
In SF networks, due to the presence of nodes with large degree, 
the method would be useful. 
We check numerically the numbers of toppling events 
and distinct nodes participating to a given avalanches, finding 
that they are scaled in a similar fashion. Thus the avalanches 
tend to form tree structures with little loops, 
supporting the validity of the branching process approach.    
We obtained the exponent of the avalanche size distribution 
$\tau=\gamma/(\gamma-1)$ and the dynamic exponent $z=(\gamma-1)/(\gamma-2)$ 
in the range $2 < \gamma < 3$, while for $\gamma > 3$, they have 
mean-field values $\tau=3/2$ and $z=2$. At $\gamma=3$, 
a logarithmic correction appears. We also performed 
numerical simulations, finding that the exponents obtained from numerical 
simulations behave similarly to the analytic solutions.  
Next we consider the case of uniform threshold height, 
obtaining that $\tau=3/2$ and $z=2$ for all $\gamma > 2$ 
analytically and numerically. 

{\it Numerical simulations}---We use  
the static model~\cite{static} to generate SF networks. 
We first start with $N$ nodes, each of which is indexed 
by an integer $i$ $(i=1,\dots,N)$ and is assigned a weight equal 
to $w_i=i^{-\alpha}$. 
Here $\alpha$ is a control parameter in $[0,1)$ and
is related to the degree exponent via the relation
$\gamma=1+1/\alpha$ for large $N$. Second, we 
select two different nodes $i$ and $j$ with probabilities 
equal to the normalized weights, $w_i/\sum_k w_k$ and 
$w_j/\sum_k w_k$, respectively, and attach an edge between 
them unless one exists already. This process is repeated until 
the mean degree of the network becomes $2m$, where we use $N=10^6$ and 
$m=2$ in this work. 

Next, we perform the dynamics on the SF network following 
the rules: 
(i) At each time step, a grain is added at a randomly chosen node $i$. 
(ii) If the height at the node $i$ reaches or exceeds a prescribed 
threshold $z_i$, where we set $z_i=k_i$, the degree of the node 
$i$, then it becomes unstable and all the grains
at the node topple to its adjacent nodes;
\begin{equation}
h_i \rightarrow h_i-k_i,~~~{\rm and}~~~h_j=h_j+1~~, 
\end{equation}
where $j$ is a neighbor of the node $i$.
During the transfer, there is a small fraction $f=10^{-4}$ of 
grains being lost, which plays the role of sinks without
which the system becomes overloaded in the end. 
(iii) If this toppling causes any of the adjacent nodes 
to be unstable, subsequent topplings follow on those nodes 
in parallel until there is no unstable node left, 
forming an avalanche. (iv) Repeat (i)--(iii).
The avalanches without loss of any grains are regarded as
``bulk'' avalanches and taken into consideration hereafter.
Note that each node has its own threshold, being equal to its degree, 
which is different from the models on regular lattices. 

After a transient period, we measure the following quantities at each 
avalanche event:
(a) the avalanche area $A$, $i.e.,$ the number of distinct nodes 
participating in a given avalanche,
(b) the avalanche size $S$, $i.e.,$ the number of toppling 
events in a given avalanche,
(c) the number of toppled grains $G$ in a given avalanche,
and (d) the duration $T$ of a given avalanche.
To obtain the probability distribution of each quantity,
we perform the statistical average over at least $10^6$ avalanches
after reaching the steady state.

\begin{figure}[t]
\centerline{\epsfxsize=8.5cm \epsfbox{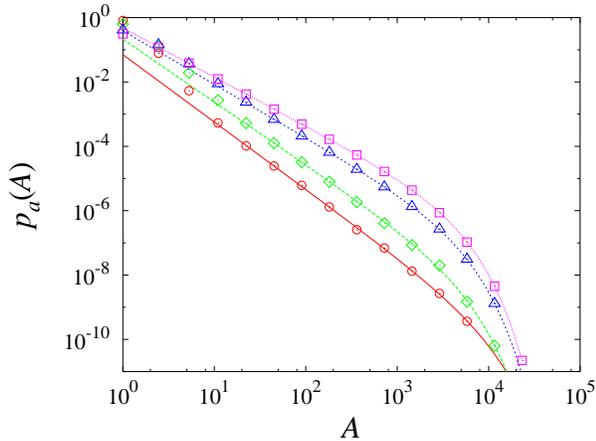}}
\caption{The avalanche size distributions for the static model
of $\gamma=\infty$ (magenta {$\Box$}), 3.0 (blue {$\triangle$}), 
2.2 (green {\Large$\diamond$}), and 2.0 (red {\Large$\circ$}).
The data are fitted with a functional form of Eq.~(1).
For the fitted values of $\tau$, see Table I. Data are logarithmically binned.}
\end{figure}

The avalanches usually do not form a loop, as
the probability distributions of the two quantities $A$ and $S$ 
behave in a similar fashion. For example, the maximum area and 
size ($A_{\rm max}$, $S_{\rm max}$) among avalanches are 
(5127, 5128), (12058, 12059) and (19692, 19692) 
for $\gamma=2.01$, 3.0 and $\infty$, respectively. 
So we shall not distinguish $A$ and $S$ but keep our attention 
mainly on the avalanche area distribution
which has the most direct implication in connection
with cascading failure phenomena in real-world networks. 
The avalanche area distribution fits well to Eq.~(1), where 
$s$ can represent either $A$ or $S$. 
In order to check, we study the case of the ER graph, which actually 
is the case of $\alpha=0$ of the static model, obtaining 
$\tau=1.52(1)$, consistent with the known result~\cite{bonabeau}. 
As $\gamma$ decreases from $\infty$, $\tau$ for $\gamma=5.0$ 
is more or less the same, but beyond $\gamma=3$, it increases 
rather noticeably with decreasing $\gamma$ in Table 1.  
Those values are compared with the ones obtained analytically 
below, showing a reasonable agreement. 
The discrepancy can be attributed to the finite-size effect.
Also the probability of losing a grain ($f=10^{-4}$) sets
a characteristic size of the avalanche, roughly as $s_c\sim
1/(2mf)$.
   
It is worthwhile to note that the case of $\tau > 1.5$ has 
never been observed in 
the Euclidean space, suggesting that the dynamics of the 
avalanche on SF networks differs from what is expected from 
the mean-field prediction. 
This feature have also been seen in other problems on SF networks
such as the ferromagnetic ordering of the Ising model \cite{ising}
and the percolation problem \cite{percol}. 

\begin{table}[t]
\begin{tabular}{cccccc}
\hline\hline
\phantom{aa}$\gamma$\phantom{aa}&\phantom{aa}$\tau_m$\phantom{aa}&
\phantom{aa}$\tau_t$\phantom{aa}&\phantom{aa}$z_m$\phantom{aa}&
\phantom{aa}$z_t$\phantom{aa}\\
\hline
$\infty$ & 1.52(1)&1.50 & 1.8 & 2.00\\
5.0 & 1.52(3) &1.50 & 1.9 & 2.00\\
3.0$^*$ & 1.66(2)&1.50 & 2.2 & 2.00\\
2.8 & 1.69(3) &1.56& 2.3 & 2.25 \\
2.6 & 1.75(4) & 1.63 & 2.5 & 2.67 \\
2.4 & 1.89(3) & 1.71& 2.8 & 3.50\\
2.2 & 1.95(9) & 1.83 & 3.5 & 6.00\\
2.01 & 2.09(8) & 2.0 & $ $ & $\infty$\\
\hline\hline
\end{tabular}
\caption{Values of the avalanche size exponent $\tau$ and the 
dynamic exponent $z$ for the static model with mean degree 4 and 
of size $N=10^6$. The subscripts $m$ and $t$ mean the measured 
and theoretical values, respectively. Note that since the dynamic 
exponent diverges theoretically as $\gamma \rightarrow 2$, 
numerical simulation data contain lots of fluctuations from 
sample to sample. $^*$The case of $\gamma=3$ has logarithmic corrections
in $\tau_t$ and $z_t$.  }
\end{table}

We have also considered the avalanche duration distribution.
Since the duration of an avalanches does not run long enough 
due to the small-world effect, the duration distribution is not well 
shaped numerically with finite size systems. 
Instead, we address this issue rather in an indirect manner. 
We measure the dynamic exponent $z$ in the relation between 
avalanche size and duration,  
\begin{equation}
s\sim t^{z}
\end{equation}
for large $t$. Numerical values of $z$ for different $\gamma$ 
are tabulated in Table I.

{\it Branching process}---
Since the quantities $A$ and $S$ scale in a similar 
manner, it would be reasonable to view the avalanche dynamics 
on SF networks as a multiplicative branching process~\cite{harris}.  
To each avalanche, one can draw a corresponding tree structure: 
The node where the avalanche is triggered is the originator of the 
tree and the branches out of that node correspond to topplings to 
the neighbors of that node. As the avalanche proceeds, the tree grows. 
The number of branches of each node on the tree is not uniform 
but it is nothing but its own degree. 
The branching process ends when no further avalanche proceeds. 
In the tree structure, a daughter-node born at time $t$ is 
located away from the originator by distance $t$ along the 
shortest pathway. In branching process, 
it is assumed that branchings from different parent-nodes occur 
independently. Then one can derive the statistics of avalanche size 
and lifetime analytically from the tree structure~\cite{otter,alstrom}. 
Note that the size and the lifetime of a tree correspond to the 
avalanche size $s$ and the avalanche duration $t$ of a single 
avalanche, respectively.

To be more specific, we introduce the probability $q_k$ that a certain 
node generates $k$ branches, which is given by  
\begin{equation}
q_k= \frac{kp_d(k)}{\sum_{j=1}^{\infty}jp_d (j)}\frac{1}{k} = 
\frac{k^{-\gamma}}{\zeta(\gamma-1)} ~~~~{\rm for}~~~ k \ge 1,
\label{eq:qk}
\end{equation}
where $\zeta(x)$ is the Riemann zeta function.
$q_0=1-\sum_{k=1}^{\infty}q_k=1-{\zeta(\gamma)}/{\zeta(\gamma-1)}$. 
In Eq.~(\ref{eq:qk}), the factor 
${kp_d(k)}/{\sum_{j=1}^{\infty}jp_d (j)}$ represents the normalized 
probability that the node gains a grain from one of its neighbors 
and $1/k$ is the probability that the node has height $k-1$ 
before toppling occurs. The factor $1/k$ comes from 
the assumption that there is no any typical height 
of a node in inactive state regardless of its degree $k$, 
and toppling can be triggered only when the height is $k-1$.
The assumption was checked numerically and holds reasonably well.
Note that with $q_k$ of Eq.~(4), the criticality condition
$\sum_{k=0}^{\infty} k q_k=1$ is automatically satisfied.

Let ${\cal P}(y)=\sum_{s=1}^\infty p(s) y^s$ and  
${\cal Q}(\omega)=\sum_{k=0}^\infty q_k \omega^k$ 
be the generating functions of a tree-size distribution 
$p(s)$ and $q_k$, respectively. Then following 
the theory of multiplicative branching process~\cite{otter,harris}, 
one finds that they are related as 
\begin{equation}
{\cal P} (\omega) = \omega {\cal Q}({\cal P} (\omega)). 
\label{eq:sc}
\end{equation}
The asymptotic behaviors of $p(s)$ can be obtained from  
the singular behavior of ${\cal Q}(\omega)$ near $\omega=1$ of 
Eq.~(\ref{eq:sc}). 

The generating function ${\cal Q}(\omega)$ is written as 
${\cal Q}(\omega)=q_0+{\rm Li}_\gamma (\omega)/\zeta(\gamma-1)$, where 
${\rm Li}_{\gamma}(\omega)$ is the polylogarithm function 
of order $\gamma$, defined as 
${\rm Li}_\gamma (\omega) = (\omega/\Gamma[\gamma]) 
\int_0^\infty (\exp(y)-\omega)^{-1} y^{\gamma-1}dy$ with 
the Gamma function $\Gamma(\gamma)$.
The polylogarithm function has a branch cut $[1,\infty)$ in the 
complex $\omega$-plane with a well-known expansion near 
$\omega=1$~\cite{robin}. As manifest in such branch-cut discontinuity, 
one can see that ${\cal Q}(\omega)$ is expanded near $\omega=1$ as
\begin{equation}
{\cal Q}(\omega)-\omega
\simeq \left\{
\begin{tabular}{ll}
$A(\gamma)(1-\omega)^{\gamma-1}$ & \phantom{a}($2<\gamma <3$), \\
$\frac{1}{\zeta(2)}\left[\frac{3}{4}+\frac{\zeta(2)-\ln (1-\omega)}{2}\right](1-\omega)^2$ & \phantom{a}($\gamma =3$), \\
$\frac{1}{2}B(\gamma)(1-\omega)^{2}$ & \phantom{a}($\gamma >3$), \\
\end{tabular}\right.
\label{eq:f_expand}
\end{equation}
to the leading order in $(1-\omega)$.
Here $A(\gamma) = \Gamma(1-\gamma)/\zeta(\gamma-1)$ and 
$B(\gamma)=\left[\zeta(\gamma-2)/\zeta(\gamma-1)\right]-1$. 
From the relation between ${\cal Q}(\omega)$ and 
${\cal P}(y)$ in Eq.~(\ref{eq:sc}), 
$\omega={\cal P}(y)$ is obtained by inverting $y=\omega/{\cal Q}(\omega)$.
The asymptotic behaviors of $p(s)$ 
for large $s$ can then be calculated through 
$2\pi i p(s)=\int_C dy {\cal P}(y) y^{-s-1}$ where $C$ is 
a contour enclosing $y=0$ but not crossing the branch-cut 
$[1,\infty)$. We obtain that 
\begin{equation}
p(s)\sim \left\{
\begin{array}{ll}
a(\gamma)\  s^{-\gamma/(\gamma-1)} & (2< \gamma <3), \\[2.5mm]
b s^{-3/2}(\ln s)^{-1/2} & (\gamma=3),\\[2.5mm]
c(\gamma)\ s^{-3/2} & (\gamma>3), 
\end{array}
\right.
\label{eq:P(s)}
\end{equation}
where 
$a(\gamma)=-A(\gamma)^{1/(1-\gamma)}/\Gamma[1/(1-\gamma)]$,
$b=\sqrt{\pi/6}$, and
$c(\gamma)=\sqrt{1/(2\pi B(\gamma))}$.
Thus the exponent $\tau$ is determined to be $\tau=\gamma/(\gamma-1)$ 
for $2<\gamma < 3$ and $\tau=3/2$ for $\gamma \ge 3$.
The value $3/2$ is consistent with the mean-field value on 
the Euclidean space. 
This behavior of $\tau$ is in reasonable agreement with that obtained 
by numerical simulations as tabulated in Table 1.
To confirm the analytic solution for $\tau$, 
we perform numerical simulations following the branching process 
of Eq.(\ref{eq:qk}). Indeed, we obtain $\tau \approx 2.0, 1.55$ and 
1.5 for $\gamma=2.01, 3.0$ and 5.0, respectively (Fig.~2).
\begin{figure}[t]
\centerline{\epsfxsize=8.5cm \epsfbox{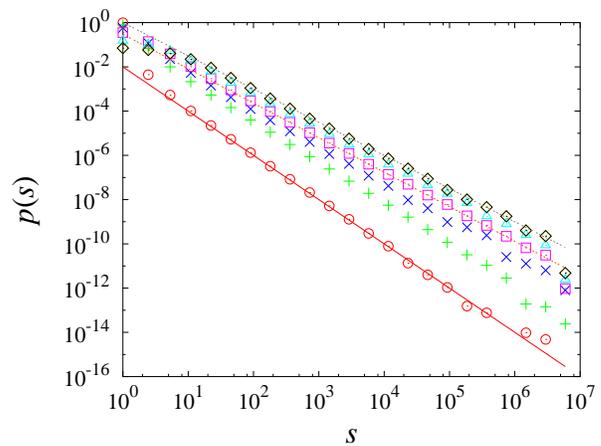}}
\caption{Simulation results of the branching process with the power-law 
branching ratio with
$\gamma=2.01$ ({\Large$\circ$}), 2.2 ($+$), 2.5 ($\times$), 3.0 ($\Box$),
4.0 ($\triangle$), and 5.0 ($\Diamond$).}
\end{figure}

The distribution of duration, $i.e.$, the lifetime of a tree growth, 
can be evaluated similarly~\cite{alstrom,harris}. 
Let $r(t)$ be the probability that a branching process stops 
at or prior to time $t$. Then it is simple to know that 
$r(t)={\cal Q}(r(t-1))$.
For large $t$, $r(t)$ comes close to $1$ and its time variation 
$r(t)-r(t-1)$ is given by 
the right hand side of Eq.~(6) with $\omega$ replaced by
$r(t-1)$ for each region of $\gamma$.
Thus the lifetime distribution $\ell(t)=r(t)-r(t-1)$ is given as
\begin{equation}
\ell(t)\sim \left\{
\begin{array}{ll}
t^{-(\gamma-1)/(\gamma-2)} ~~~~~~~& ~~~ (2< \gamma <3), \\[2.5mm]
t^{-2}(\ln t)^{-1} & ~~~ (\gamma=3), \\[2.5mm]
t^{-2}  &  ~~~ (\gamma>3).
\end{array}
\right.
\label{eq:D(t)}
\end{equation}
Since the distributions of $p(s)$ and $\ell(t)$ are originated from 
the same tree structures, $p(s){\rm d}s \sim \ell(t){\rm d}t$.   
Thus from Eqs.~(\ref{eq:P(s)}) and (\ref{eq:D(t)}), we obtain 
the dynamic exponent defined via $s\sim t^z$ as 
$z=(\gamma-1)/(\gamma-2)$ for $2<\gamma<3$ and $z=2$ for $\gamma\ge3$. 
Following the same steps, we can obtain the exponents $\tau$ 
and $z$ for more general case where $z_i= k_i^{\beta}$ with 
$0 \le \beta \le 1$.
We find $\tau=(\gamma+2\beta-2)/(\gamma+\beta-2)$ and 
$z=(\gamma+\beta-2)/(\gamma-2)$ for $2<\gamma<\gamma_c 
\equiv \beta+2$, and $\tau=1.5$ and $z=2.0$ 
for $\gamma>\gamma_c$.

\begin{figure}
\centerline{\epsfxsize=8.5cm \epsfbox{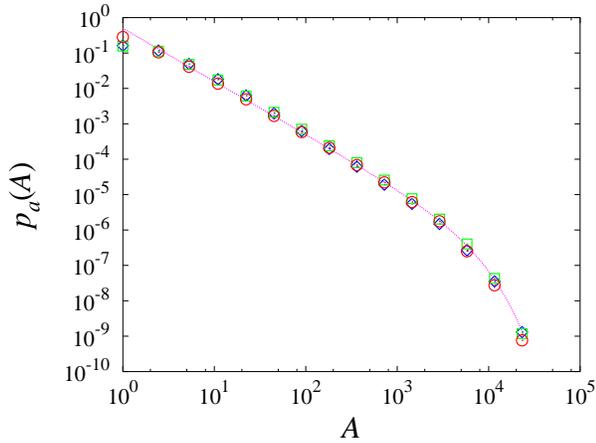}}
\caption{The avalanche size distribution for the static model
with uniform threshold. Shown are the case of 
$\gamma=\infty$ ({\Large$\circ$}), 3.0 ({\Large$\diamond$}), 
and 2.2 ($\Box$).  
All estimated values of $\tau$ are $1.50\pm0.05$.
}
\label{manna}
\end{figure}

{\it Sandpile with uniform threshold}---
We also consider the case that the threshold height of each node is 
uniform, while its degree is distributed following the power law. 
To realize this, we choose the threshold to be $z_i=2$ for vertices of 
degree larger than 1, and $z_i=1$ for those of degree 1.
Then we modify the dynamic rule accordingly: 
Toppled grains are transferred to $z_i$ randomly selected nearest 
neighbors, which is similar to that of the Manna model~\cite{manna}.
In this case, we obtained $\tau\simeq1.5$ in all $\gamma>2$ (Fig.~\ref{manna}).
This can easily be understood through the branching process analogy:
In this case, we have $q_1 = p_d(1)/\langle k\rangle$, 
$q_0=q_2 = \sum_{k=2}^{\infty} \frac{kp_d(k)}{\langle k\rangle}\frac{1}{2} = 
\frac{1-q_1}{2}$, and $q_k=0$ for $k>2$.
Thus ${\cal Q}(\omega)$ is analytic for all $\omega$, yielding the usual
mean-field exponents $\tau=3/2$ and $z=2$ for all $\gamma>2$.  

{\em Summary}---
We have studied the Bak-Tang-Wiesenfeld sandpile model
on scale-free networks. To account for the high heterogeneity of the system,
the threshold height of each vertex is set to be 
the degree of the vertex. 
Numerical simulations suggest that for $2<\gamma< 3$ 
the scaling behavior
of the avalanche size distribution differs from the 
mean-field prediction.
By mapping to the multiplicative branching process,
we could obtain the asymptotic behaviors of the avalanche size 
and duration distributions analytically.
They are described by novel exponents $\tau$ and $z$
different from the simple mean-field predictions.
The result remains the same when threshold contains noise as 
$z_i=\eta_i k_i$ with $\eta_i$ being distributed 
uniformly in [0,1] and when a new grain is added to a node chosen 
with probability proportional to the degree of that node. 
In the case of uniform threshold, on the other hand, 
we get the mean-field behaviors for all $\gamma>2$.

The fact that $\tau$ increases as $\gamma$ decreases implies
the resilience of the network under avalanche phenomena,
by the role of the hubs that sustain large amount of grains 
thus playing the role of a reservoir.
This is reminiscent of the extreme resilience of the network 
under random removal of vertices for $\gamma\le3$ \cite{bara,cohen1,newman1}. 
While preparing this manuscript, we have learned of a recent work
by Saichev {\it et al.} motivated by the study of earthquake avalanches \cite{sornette}.
Their results partly overlap with ours.

\begin{acknowledgments}
This work is supported by the KOSEF Grant No. R14-2002-059-01000-0 
in the ABRL program. 
\end{acknowledgments}

\end{document}